\documentclass[11pt]{article}
\usepackage{epsfig}
\newcommand{\bM} {{\mathbf M}}
\newcommand{\bK} {{\mathbf K}}
\newcommand{\bN} {{\mathbf N}}
\newcommand{\bL} {{\mathbf L}}
\newcommand{\bA} {{\mathbf A}}

\newcommand{\bS} {{\mathbf S}}

\newcommand{\bgrad} {\bm{\nabla}}
\newcommand{\be}{\begin{equation}}
\newcommand{\ee}{\end{equation}}
\newtheorem{defn}{Definition}
\newcommand{\bed}{\begin{defn}}
\newcommand{\eed}{\end{defn}}
\newcommand{\bea}{\begin{eqnarray}}
\newcommand{\eea}{\end{eqnarray}}
\newcommand{\beas}{\begin{eqnarray*}}
\newcommand{\eeas}{\end{eqnarray*}}
\newcommand{\ba}{\begin{array}}
\newcommand{\ea}{\end{array}}

\newcommand{\bC}{\hbox{{\rm I}
\kern-.8em\hbox{{\rm C}}}}
\newcommand{\bR}{\hbox{{\rm I}\kern-.2em\hbox{{\rm R}}}}
\newcommand{\bm}[1]{\mbox{\boldmath $#1$}}
\newcommand{\bc}{\begin{center}}
\newcommand{\ec}{\end{center}}
\newcommand{\half}{\mbox{$\frac{1}{2}$}}
\newcommand{\rf}[1] {(\ref{#1})}
\newcommand{\dd}[2] {\frac{\partial{#1}}{\partial{#2}}}
\newcommand{\mb}[1]{\mathbf {#1}}
\newcommand{\nn}{\nonumber}
\def\C{{\rm\kern.24em \vrule width.02em height1.4ex depth-.05ex \kern-.26emC}}
\def\R{{\rm I\kern-.20em R}}
\def\Z{{\rm\kern.26em \vrule width.02em height0.5ex depth0ex \kern.04em
\vrule  width.02em height1.47ex depth-1ex \kern-.34em Z}}
\def\N{{\rm I\kern-.20em N}}
\def\Q{{\rm\kern.24em \vrule width.02em height1.4ex depth-.05ex \kern-.26emQ}}

\begin{document}

\title{Backward error analysis for  multisymplectic discretizations of 
Hamiltonian PDEs}
\author{A.L. Islas\thanks{Department of Mathematics,
University of Central Florida, aislas@mail.ucf.edu} \,
and C.M. Schober\thanks{Department of Mathematics,
University of Central Florida, cschober@mail.ucf.edu}}

\date{}

\maketitle

\begin{abstract}
Several recently developed multisymplectic schemes for Hamiltonian
PDEs have been shown to preserve  associated local conservation laws
and constraints very well in long time numerical simulations. 
Backward error analysis for PDEs, or the method of modified equations,
 is a useful technique for studying the 
qualitative behavior of a discretization and  provides insight 
into the preservation properties of the scheme.
In this paper we initiate a backward error analysis for PDE
discretizations, in particular of multisymplectic  box schemes 
for the nonlinear Schrodinger equation. We show that the 
associated modified differential equations 
are also multisymplectic and
derive the  modified conservation laws which 
are satisfied
to higher order by the numerical solution.
Higher order preservation of the modified local conservation laws 
is verified numerically.
\end{abstract}

 
\section{Introduction}
When developing numerical integrators for Hamiltonian PDEs that 
possess a multisymplectic structure
(i.e. symplectic in both space and time),
it is natural to require the
numerical scheme to preserve exactly a discrete version of 
the multisymplectic conservation law (MSCL) \cite{br2,mars2}.
However, this  does not 
imply preservation of 
other dynamical invariants of the system such as the local energy and 
momentum conservation laws or 
global invariants which determine the phase space structure.
A question that immediately arises then is, to what extent are 
the other invariants of the system preserved?
Recent numerical experiments using multisymplectic integrators
for nonlinear 
wave equations (e.g. the nonlinear Schrodinger (NLS), sine-Gordon,
and Gross-Pitaevskii eqautions)
show that the local conservations laws are preserved 
very well, although not exactly, over long 
times \cite{iskasc,issc03,issc03b}.
Further, the improved preservation of the local conservation laws
is reflected 
in an improved preservation of complicated phase space structures
\cite{issc03b}.
This is reminiscent of the behavior of  symplectic schemes 
for Hamiltonian ODEs. 
Symplectic integrators are  designed to preserve the symplectic 
structure, not to preserve the energy.  In fact, for general Hamiltonian
systems, conservation of the symplectic structure and conservation of 
energy are conflicting requirements that, in general, are not solved 
simultaneously by a given scheme \cite{htext}. 
Even so,  symplectic integrators preserve the Hamiltonian 
extremely well over  very long times.

Backward error analysis (BEA), or the method of modified equations,
is a particularly insightful technique for studying the 
qualitative behavior of a discretization as well as an alternative 
method for checking the accuracy of the numerical solution
\cite{pdetext}. 
Since our main interest lies in the 
geometry preserving properties of multisymplectic schemes,
the main question backward error analysis tries to answer (whether
the distinguishing  properties of the original equation
carry over to the modified equation which the numerical solution
satisfies to higher order) becomes
relevant to our study.
For a given scheme, the derivation of the associated 
modified equation is related to the calculation of 
the local truncation error and 
has, typically, been used to examine the dispersive, dissipative 
and diffusive properties of PDE discretizations. 
For example, in the numerical analysis of linear PDEs
a backward error analysis of the Lax-Friedrichs method
or the upwind method 
for the advection equation produces in both cases 
a modified equation that is an
advection-diffusion equation. This helps one to understand
the qualitative behavior of the methods and,
from this perspective, explains why the numerical solution 
in both cases becomes smeared out as time evolves.

Likewise, BEA is an important tool in the study of geometric
integrators \cite{hlu97,htext,moore03,moore03b}. 
For Hamiltonian ODEs, symplectic methods lead to modified equations
which are also Hamiltonian. In fact, the modified equation of a Hamiltonian 
ODE is also Hamiltonian if and only if the integrator is symplectic;
this is then  used to rigorously establish
that a symplectic integrator almost preserves the total energy
over an exponentially long period of time \cite{htext}.
In striking contrast, nonsymplectic methods used to integrate  Hamiltonian 
ODEs can introduce dissipation, a feature which is readily predicted by
the dissipative form of the modified equations.
Less has been established using BEA for Hamiltonian PDEs since there are
a variety of  ways to implement a BEA and the relevance of the 
analysis is open to interpretation.
Spatial discretization of a PDE results in a system of ODES to
which a standard BEA can be applied to derive a modified equation that
is satisfied to higher order in one independent variable.
Alternatively, a BEA can be used to derive  modified equations 
for the PDE that are satisfied to 
higher order in both space and time \cite{moore03,moore03b}.

In this paper 
we implement a formal backward error analysis in both space and time 
of two  multisymplectic box schemes,
the Euler and the centered cell box schemes,
as applied to the nonlinear Schrodinger equation.
We find that the modified equations of 
these box schemes are also multisymplectic. 
The modified  PDEs  are used to derive modified conservation laws 
of energy and momentum that are 
approximated by the MS scheme to higher order in space and time.
For the 
centered cell discretization of the NLS we numerically verify
that the modified conservation
laws are satisfied to higher order by the numerical solution.
This provides a partial 
explanation of the superior resolution  of the local 
conservation laws and global invariants by  MS schemes 
(e.g. see the numerical experiments in section 5) 
and a deeper understanding of the
local and global properties of MS integrators.

The paper is organized as follows. In the next section  we recall 
the multisymplectic formulation of Hamiltonian  PDEs and 
of the NLS equation. 
In section 3 we introduce the box schemes, establish multisymplecticity,
and apply them to the NLS equation.
We present a straightforward  method for 
obtaining compact box schemes that is applicable to  many
multisymplectic PDEs.
Section 4 contains the backward error analysis 
of the discretizations.
In section 5 numerical experiments for the 
MS centered cell box scheme are discussed, 
illustrating the remarkable behavior of MS schemes.
Higher order preservation of the modified local conservation laws 
is verified numerically, which 
supports the use of MS
integrators in long time numerical simulations of Hamiltonian PDEs.

\section{Multisymplectic Hamiltonian PDEs}
A Hamiltonian PDE (in the ``1+1'' case) 
is said to be  
multisymplectic  if it can be written as
\be
\bM z_t + \bK  z_x = 
\bgrad_z S,\qquad z\in\R^n,
\label{multi}
\ee
where $\bM,\, \bK \in \R^{n\times n}$ are skew-symmetric matrices
and $S:\,\R^n \rightarrow \R$ 
is a smooth function of the state
variable $z(x,t)$ \cite{reich1,br2}.
The variational equation associated with  (\ref{multi})  is  
given by
\be
\bM dz_t +  \bK \, dz_{x} = {\mathbf S}_{zz} dz.
\label{var}
\ee
The Hamiltonian system \rf{multi} is multisymplectic in the sense that 
associated with $\bM $ and $\bK $ are the  2-forms
\be
\omega = \frac{1}{2} (dz\wedge \bM  dz),\qquad
\kappa = \frac{1}{2} (dz\wedge \bK  dz),
\label{multi_twoforms}
\ee
which define a symplectic space-time 
structure (symplectic with respect to more than one independent
variable).

Any system of the form \rf{multi} satisfies conservation of symplecticity.
Let $dz$ be any solution of the variational equation \rf{var}.
Then it can be shown that 
$\omega$ and $\kappa$, as defined in \rf{multi_twoforms},
satisfy the multisymplectic conservation law (MSCL):
\be
\label{multi_scl}
\dd{\omega}{t} + \dd{\kappa}{x} = 0.
\ee
This result is obtained by noting that  
\bea
2 \omega_t = \left({dz\wedge\bM  dz}\right)_t
&=& dz_t\wedge  \bM  dz + dz\wedge  \bM  dz_t\nn\\
&=& -\left(\bM  dz_t\right)\wedge   dz 
+ dz\wedge  \bM  dz_t\nn\\
&=& -\left(\bS_{zz}dz - \bK  dz_x\right)\wedge   dz 
+ dz\wedge  \left(\bS_{zz}dz - \bK  dz_x\right)\nn\\
&=& - \left( dz_x\wedge  \bK  dz + dz\wedge  \bK  dz_x\right)\nn\\
&=& - \left( dz\wedge  \bK  dz \right)_x = - 2 \kappa_x\nn
\eea
since $\bM ,\,\bK $ are skew-symmetric and $\bS_{zz}$ is
symmetric.
The MSCL \rf{multi_scl} is a local property and 
expresses the fact that  symplecticity for Hamiltonian PDEs can
vary locally over the spatial domain.

An important consequence of the MS structure is that when the Hamiltonian
$S(z)$ is independent of $t$ and $x$, the PDE has {\em local} energy
and momentum conservation laws \cite{reich1,br2}
\bea
E_t + F_x &=& 0,\qquad
E = S(z) + \half z_x^T\bK \, z,\qquad
F = -\half z^T_t\bK \, z,
\label{ecl} \\
I_t + G_x &=& 0,\qquad
G = S(z) + \half z_t^T\bM \, z,\qquad
I = -\half z^T_x\bM \, z.
\label{mcl}
\eea
For periodic boundary conditions, the local conservation laws 
can be integrated in $x$ to obtain  global conservation of energy
and momentum. 

\subsection{Multisymplectic formulation of  the NLS  equation}
The focusing one dimensional nonlinear Schr\"odinger (NLS) equation,
\be
iu_t + u_{xx} + 2|u|^2 u = 0,
\label{NLS}
\ee
can be written in multisymplectic form by letting 
$u=p+iq$ and introducing the new variables $v = p_x,\,w=q_x$.
Separating \rf{NLS} into real and imaginary parts, 
we obtain the system \cite{iskasc}:
\be
\ba{rcl}
q_t - v_x &=& 2\left(p^2+q^2\right)p\\
-p_t - w_x &=& 2\left(p^2+q^2\right)q\\
p_x &=& v\\
q_x &=& w,
\ea
\label{nls}
\ee
which is equivalent to
the multisymplectic form \rf{multi}
for the NLS equation with 
\[
z = \left(\ba{c}
p\\q\\v\\w
\ea \right),\qquad
\bM  = \left(\ba{cccc}
0 & 1 & 0 & 0 \\ -1 & 0 & 0 & 0 \\ 0 & 0 & 0 & 0 \\ 0 & 0 & 0 & 0
\ea \right),
\qquad \bK  = \left(\ba{cccc}
0 & 0 & -1 & 0 \\ 0 & 0 & 0 & -1 \\ 1 & 0 & 0 & 0 \\ 0 & 1 & 0 & 0
\ea \right),
\]
and Hamiltonian 
\[
S = \frac{1}{2}\left[\left(p^2 + q^2\right)^2 +
v^2 + w^2\right].
\]

Implementing \rf{ecl}-\rf{mcl} for the NLS 
equation yields
the local energy conservation law (LECL) 
\be
E_t + F_x = 0,\qquad
E = \frac{1}{2}\left[\left(p^2+q^2\right)^2 
- v^2 - w^2\right],\quad
F = vp_t + wq_t,
\label{nls_ecl}
\ee
and the local momentum conservation law (LMCL) 
\be
I_t + G_x = 0,\qquad
I = pw - qv,\quad
G = \left(p^2+q^2\right)^2+v^2+w^2
-\left(pq_t - p_t q\right).
\label{nls_mcl}
\ee
Additionally we have a
norm conservation law for the NLS equation
\be
N_t + M_x = 0,\qquad
N = \frac{1}{2}\left(p^2 + q^2\right),\quad
M = qv - pw.
\label{nls_ncl}
\ee
These three equations, when integrated with respect to $x$, yield 
the classic
global conservation of energy ${\cal E}(t)$ (Hamiltonian), 
momentum ${\cal I}(t)$ and norm ${\cal N}(t)$.



\section{Multisymplectic box schemes}
Multisymplectic discretizations are numerical schemes for approximating 
\rf{multi} which preserve a discrete version of the multisymplectic
conservation law \rf{multi_scl}.
That is, if
the discretization of the multisymplectic PDE 
and its  conservation law are written schematically as
\be
\bM \partial_t^{i,j} z_i^j + \bK \partial_x^{i,j} z_i^j = 
\left(\bgrad_{z} S(z_i^j)\right)_i^j,
\label{dmulti}
\ee
and 
\be
\partial_t^{i,j}\omega_i^j + \partial_x^{i,j}\kappa_i^j = 0,
\label{dscl}
\ee
where $f_i^j = f(x_i,t_j)$,
and $\partial_t^{i,j}$ and $\partial_x^{i,j}$ are discretizations of the 
corresponding derivatives $\partial_t$ and $\partial_x$,
then the numerical scheme \rf{dmulti} is said to be  multisymplectic 
if \rf{dscl} is a  discrete  conservation law
of \rf{dmulti} \cite{reich1,br2}.

A standard method for constructing multisymplectic schemes is to apply
a known symplectic discretization to each independent variable.
For example, splitting the matrices $\bM$ and $\bK$ as
\be
\bM = \bM_+ + \bM_- \quad\mbox{and}\quad \bK = \bK_+ + \bK_-
\quad\mbox{with}\quad
\bM_+^T = -\bM_- \quad\mbox{and}\quad \bK_+^T = - \bK_-,
\label{split}
\ee
and using the symplectic Euler forward-backward difference
approximations on both space and time derivatives yields the Euler box scheme
\be
\bM_+\frac{z_0^1 - z_0^0}{\Delta t} +
\bM_-\frac{z_0^0 - z_0^{-1}}{\Delta t} +
\bK_+\frac{z_1^0 - z_0^0}{\Delta t} +
\bK_-\frac{z_0^0 - z_0^{-1}}{\Delta t} = 
\bgrad_z S(z_0^0).
\label{Ebox}
\ee

Similarly, applying the symplectic midpoint rule 
to both the time and space derivatives in \rf{multi}
yields a ``centered cell'' box discretization
\be
\bM \left(\frac{z_{1/2}^{1} - z_{1/2}^0}{\Delta t}\right)
+ \bK \left(\frac{z_{1}^{1/2} - z_0^{1/2}}{\Delta x}\right)
=\bgrad_z S\left(z_{1/2}^{1/2}\right),
\label{CCbox}
\ee
where
\be
z_{1/2}^j = \half\,\left(z_0^j + z_1^j\right),\;
z_i^{1/2} = \half\,\left(z_i^0 + z_i^1\right),\;
z_{1/2}^{1/2} = \frac{1}{4}
\left(z_0^0 + z_0^1 + z_1^0 + z_1^1\right).
\label{notation}
\ee
The local truncation error for the Euler box scheme is 
${\cal O}\left(\Delta t + \Delta x^2\right)$,
while for the centered cell discretization it is 
${\cal O}\left(\Delta t^2 + \Delta x^2\right)$.

Multisymplecticity of schemes \rf{Ebox} and \rf{CCbox} is easily established.
For example, to do so for the centered cell scheme, we use the 
discrete variational equation associated with \rf{CCbox} given by
\be
\bM \left(\frac{dz_{1/2}^{1} - dz_{1/2}^0}{\Delta t}\right)
+ \bK \left(\frac{dz_{1}^{1/2} - dz_0^{1/2}}{\Delta x}\right)
=\bS_{zz}\,dz_{1/2}^{1/2}.
\label{multi_mpr_ve}
\ee
Taking the wedge product of $dz_{1/2}^{1/2}$ with \rf{multi_mpr_ve}, 
note that the right-hand side is zero, 
since \(\bS_{zz}\) is symmetric. The terms 
on the left-hand side can be simplified
\bea
dz_{1/2}^{1/2} \wedge \bM \left(dz_{1/2}^1 - dz_{1/2}^0\right) &=&
\half\,\left(dz_{1/2}^1 + dz_{1/2}^0\right)\wedge 
\bM \left(dz_{1/2}^1 - dz_{1/2}^0\right)\nn\\
&=&\half\,\left(dz_{1/2}^1\wedge \bM  dz_{1/2}^1 -
dz_{1/2}^0\wedge \bM  dz_{1/2}^0\right)\nn\\
&=& \omega_{1/2}^1 -  \omega_{1/2}^0,\nn 
\eea
whereas,
\bea
dz_{1/2}^{1/2} \wedge \bK \left(dz_1^{1/2} - dz_0^{1/2}\right) &=&
\half\,\left(dz_1^{1/2} + dz_0^{1/2}\right)\wedge 
\bK \left(dz_1^{1/2} - dz_0^{1/2}\right)\nn\\
&=&\half\,\left(dz_1^{1/2}\wedge \bK  dz_1^{1/2} -
dz_0^{1/2}\wedge \bK  dz_0^{1/2}\right)\nn\\
&=& \kappa_1^{1/2} -  \kappa_0^{1/2}.\nn 
\eea
This implies that the numerical scheme \rf{CCbox}
satisfies the discrete multisymplectic conservation
law
\[
\left(\frac{\omega_{1/2}^1 - 
\omega_{1/2}^0}{\Delta t}\right)
+
\left(\frac{\kappa_1^{1/2} - 
\kappa_0^{1/2}}{\Delta x}\right)
= 0.
\]

\subsection{Multisymplectic box schemes for the  NLS equation}
The multisymplectic centered cell box scheme was first developed for the NLS 
equation in \cite{iskasc} where an apparently ad hoc reduction 
provided  a particularly compact form of the scheme.
This reduction turns out to be generalizable as can be seen 
in McLachlan's derivation of box schemes for the Korteweg 
de Vries equation \cite{mcl03}. Here  we present a
general approach for constructing compact box
schemes which is applicable to many 
multisymplectic PDEs.

\subsubsection{Euler box scheme for the NLS equation}
We begin by  introducing the following finite difference operators
\[
D_t^\pm z = \pm \frac{z_i^{\pm 1} - z_i^0}{\Delta t} \quad\mbox{and}\quad 
D_x^\pm z = \pm \frac{z_{\pm 1}^j - z_0^j}{\Delta x}.
\]
In terms of these operators the Euler box scheme \rf{Ebox} takes the form
\be
\bM_+D_t^+ z +
\bM_-D_t^- z +
\bK_+D_x^+ z +
\bK_-D_x^- z =
\bgrad_z S(z_0^0).
\label{Eboxdiff}
\ee
For the NLS,   $\bM$ and $\bK$ are split  using \rf{split}, where 
\[
\bM_+ = \left(\ba{cccc}
0 & 1 & 0 & 0\\
0 & 0 & 0 & 0 \\
0 & 0 & 0 & 0 \\
0 & 0 & 0 & 0
\ea\right)
\qquad
\mbox{ and }
\qquad
\bK_+ = \left(\ba{cccc}
0 & 0 & -1 & 0\\
0 & 0 & 0 & 0 \\
0 & 0 & 0 & 0 \\
0 & 1 & 0 & 0
\ea\right).
\]
Applying \rf{Eboxdiff}  to the NLS system \rf{nls} yields the system
\[
\ba{rcl}
D_t^+ q - D_x^+ v &=& 2\left(p^2 + q^2\right) p\\
-D_t^- p - D_x^- w &=& 2\left(p^2 + q^2\right) q\\
D_x^- p &=& v\\
D_x^+ q &=& w.
\ea
\]
After eliminating $v$ and $w$ the system reduces to
\[
\ba{rcl}
D_t^+ q - D_x^2 p &=& 2\left(p^2 + q^2\right) p\\
-D_t^- p - D_x^2 q &=& 2\left(p^2 + q^2\right) q,
\ea
\]
where we have set $D_x^2 = D_x^+ D_x^- = D_x^- D_x^+$.
When the second equation is shifted in time, 
the resulting  six-point box scheme in stencil format is :
\bea
\frac{1}{\Delta t}\left[\ba{r}
1\\ -1
\ea\right]
q - \frac{1}{\Delta x^2}\left[\ba{ccc}
0 & 0 & 0\\ 1 & -2 & 1\ea\right]p 
&=& \left[\ba{ccc}
0 & 0 & 0\\ 0 & 1 & 0\ea\right]
2\left(p^2 + q^2\right) p\nn\\
\frac{1}{\Delta t}\left[\ba{r}
1\\ -1
\ea\right]
p - \frac{1}{\Delta x^2}\left[\ba{ccc}
1 & -2 & 1\\
0 & 0 & 0\ea\right]q 
&=& \left[\ba{ccc}
0 & 1 & 0\\ 0 & 0 & 0\ea\right]
2\left(p^2 + q^2\right) q.\nn
\eea

\subsubsection{Centered cell box scheme for the NLS equation}
As before, we begin by introducing  the appropriate finite difference operators
\be
M_t z = \frac{z_i^0 + z_i^1}{2},\quad M_x z = \frac{z_0^j + z_1^j}{2},\quad
D_t z = \frac{z_i^1 - z_i^0}{\Delta t},\quad
D_x z = \frac{z_1^j - z_0^j}{\Delta x}.
\label{oprs}
\ee
In terms of these operators, the centered-cell discretization
\rf{CCbox} becomes
\be
\bM  D_t M_x z
+ \bK  D_x M_t z
=\bgrad_z S\left(M_x M_t z \right),
\label{mpr_opr}
\ee
with discrete conservation law
\[
dz\wedge \bM D_t M_x dz + dz\wedge \bK D_x M_t dz = 0.
\]

The system which results upon applying  \rf{mpr_opr} 
to \rf{nls} is
\be
\ba{rcl}
D_t M_x q - D_x M_t v &=& 2\left[\left(M_xM_t p\right)^2 +
\left(M_xM_t q\right)^2\right] M_xM_t p\\
- D_t M_x p - D_x M_t w &=& 2\left[\left(M_xM_t p\right)^2 +
\left(M_xM_t q\right)^2\right] M_xM_t q\\
D_x M_t p &=& M_xM_t v\\
D_x M_t q &=& M_xM_t w.
\ea
\label{nls_opr}
\ee
Since the operators in \rf{oprs} commute,
by multiplying the first two equations in \rf{nls_opr} by $M_x$ 
and back substituting  $v$ and $w$ into the first two equations 
we obtain
\bea
D_t M^2_x q - D^2_x M_t p &=& 2M_x\left(\left[\left(M_xM_t p\right)^2 +
\left(M_xM_t q\right)^2\right] M_xM_t p\right),\nn\\
- D_t M^2_x p - D^2_x M_t q &=& 2M_x\left(\left[\left(M_xM_t p\right)^2 +
\left(M_xM_t q\right)^2\right] M_xM_t q\right).\nn
\eea
Recombining these equations into a single complex equation (with $u=p+iq$)
yields the multisymplectic box scheme for the NLS equation
\be
iD_tM_x^2 u + D_x^2M_t u 
-2M_x\left(\left|M_xM_t u\right|^2 M_xM_t u\right) = 0,
\label{nls_opr2}
\ee
or equivalently
\be
i\frac{u_{-1/2}^{1}+u_{1/2}^{1} -
u_{-1/2}^0 - u_{1/2}^0}{2\Delta t}
+\frac{u_{-1}^{1/2} - 2u_0^{1/2} + u_{1}^{1/2}}{\Delta x^2}
-\left(\left|u_{-1/2}^{1/2}\right|^2 u_{-1/2}^{1/2}
+ \left|u_{1/2}^{1/2}\right|^2 u_{1/2}^{1/2}\right) = 0. \label{ms-cc}
\ee
In finite difference stencil format the six-point box scheme is given by 
\bea
&&\frac{i}{\Delta t}\left[\ba{rrr}
1 & 2 & 1\\
-1 & -2 & -1
\ea\right]
u +
\frac{2}{\Delta x^2}\left[\ba{ccc}
1 & -2 & 1\\
1 & -2 & 1
\ea\right]
u \nn\\
&=& \frac{1}{32}\left(\left|\left[\ba{ccc}
1 & 1 & 0\\
1 & 1 & 0
\ea\right]
u\right|^2
\left[\ba{ccc}
1 & 1 & 0\\
1 & 1 & 0
\ea\right]u +
\left|\left[\ba{ccc}
0 & 1 & 1\\
0 & 1 & 1
\ea\right]
u\right|^2
\left[\ba{ccc}
0 & 1 & 1\\
0 & 1 & 1
\ea\right]u\right).\nn
\eea
The centered cell scheme naturally gives a two time level stencil for the NLS
equation.
If every term in \rf{nls_opr2} contained a common factor, e.g. $M_x$ or $M_t$,
further compactification would be possible. As it is,
an additional reduction of \rf{nls_opr2} is not possible.

\section{Backward Error Analysis}
A useful method for analysing the qualitative behavior
of symplectic methods for  ODEs has been
backward error analysis, 
where 
one interprets the numerical solution as the ``nearly'' exact solution of a 
modified Hamiltonian differential equation.
In this section we implement a BEA in space and time 
for the multisymplectic box schemes.
The modified differential equations 
are also multisymplectic and satisfy modified conservation laws.

\subsection{BEA for the Euler box scheme}
Let $z$ be a differentiable function that,
when evaluated at the lattice points, satisfies the Euler box
scheme \rf{Eboxdiff}. Using the Taylor series expansions in $t$
about $z = z(x_i, t_j)$ 
\[
z_i ^{j\pm 1} = z \pm \Delta t\, z_t + \half\Delta t^2\, z_{tt} \pm\cdots
\]
and equivalent expansions in $x$, 
we obtain to first order the following modified equation
\be
\bM z_t + \half\Delta t\,\left(\bM_+ - \bM_-\right)z_{tt} +
\bK z_x + \half\Delta x\,\left(\bK_+ - \bK_-\right)z_{xx} = 
\bgrad_z S(z).
\label{Meuler}
\ee
If we introduce the new matrices
\[
\bN = \half\left(\bM_+ - \bM_-\right),\quad\mbox{ and }\quad
\bL = \half\left(\bK_+ - \bK_-\right),
\]
equation \rf{Meuler} can also be written in the multisymplectic form
\[
\tilde{\bM}\tilde{z}_t + \tilde{\bK}\tilde{z}_x =
\bgrad_{\tilde{z}} \tilde{S}(\tilde{z}),
\]
where
\[
\tilde{z} = \left(\ba{c}
z\\z_t\\z_x
\ea\right),
\qquad
\tilde{\bM} = \left(\ba{ccc}
\bM & \Delta t\,\bN & \mb{0}\\
- \Delta t\,\bN  & \mb{0}  & \mb{0}\\
 \mb{0} & \mb{0} & \mb{0}
\ea\right),
\qquad
\tilde{\bK} = \left(\ba{ccc}
\bK & \mb{0} &  \Delta x\,\bL \\
 \mb{0} & \mb{0} & \mb{0}\\
- \Delta x\,\bL & \mb{0}  & \mb{0}
\ea\right),
\]
and
\[
\tilde{S}(\tilde{z}) = S -  \half\Delta t\, z_t^T\bN z_t
 -  \half\Delta x\, z_x^T\bL z_x.
\]

Applying equation \rf{Meuler} to the NLS system and eliminating $v$ and $w$ yields the reduced system
\[
\ba{rcl}
q_t + \half\Delta t\, q_{tt} - p_{xx} + \frac{1}{4}\Delta x^2\, p_{xxxx} &=& 2\left(p^2 + q^2\right) p\\
-p_t + \half\Delta t \,p_{tt} - q_{xx} + \frac{1}{4}\Delta x^2\, q_{xxxx} &=& 2\left(p^2 + q^2\right) q,
\ea
\]
or setting $u = p+iq$, the single equation
\[
iu_t + u_{xx} +  2|u|^2 u + \half\Delta t\, u_{tt} + \frac{1}{4}\Delta x^2\, u_{xxxx} = 0,
\]
which is an ${\cal O}\left(\Delta t + \Delta x^2\right)$ perturbation of the NLS.

\subsection{BEA for the centered cell box scheme}
We now  assume $z$ is a sufficiently smooth 
function that, when evaluated at the lattice points,
is a solution to the centered cell scheme \rf{mpr_opr}.
Expanding $z$ in a Taylor series 
about the midpoints $(x_{i+1/2},t_{j+1/2})$ we obtain
\bea
z_{1/2}^1 &=& z + \frac{\Delta t}{2} z_t + \half\left(\frac{\Delta t}{2}\right)^2 z_{tt} +
\frac{1}{6}\left(\frac{\Delta t}{2}\right)^3 z_{ttt}  + \cdots\nn\\
z_{1/2}^0 &=& z - \frac{\Delta t}{2} z_t + \half\left(\frac{\Delta t}{2}\right)^2 z_{tt} -
\frac{1}{6}\left(\frac{\Delta t}{2}\right)^3 z_{ttt} + \cdots ,\nn
\eea
where to simplify the notation 
 $0$ and $1$ denote the grid points, $1/2$ denotes the midpoints, and 
$z = z(x_{1/2},t_{1/2})$.
The symplectic midpoint rule approximation of the time derivative is given by
\[
\frac{z_{1/2}^1 - z_{1/2}^0}{\Delta t} = z_t + \frac{\Delta t^2}{24} z_{ttt} + {\cal O}(\Delta t^4).
\]
and, similarly, the space derivative is approximated by
\[
\frac{z_1^{1/2} - z_0^{1/2}}{\Delta x} = z_x + \frac{\Delta x^2}{24} z_{xxx} + {\cal O}(\Delta x^4).
\]
Substituting  these expansions  into (\ref{CCbox}), one finds that, to order
${\cal O}(\Delta t^4 + \Delta x^4)$,  $z$ satisfies  the modified PDE
\be
\bM z_t + \frac{\Delta t^2}{24}\bM z_{ttt}+\bK z_x 
+ \frac{\Delta x^2}{24}\bK z_{xxx}
=\bgrad_z S(z), \label{MEQN}
\ee
where all quantities are evaluated at the midpoint 
$z = z(x_{1/2},t_{1/2})$.

When applying equation \rf{MEQN} to the NLS example, the resulting
modified system of equations can be  reduced to
\[
iu_t + u_{xx} + 2|u|^2 u = - i\frac{\Delta x^2}{24} u_{ttt} - \frac{\Delta x^2}{12} u_{xxxx}.
\]
which is an ${\cal O}(\Delta t^2 + \Delta x^2)$ perturbation
of NLS.

The modified local conservation laws can be
obtained directly from equation \rf{MEQN} by multiplying the equation from the
left by $z_t$ to obtain an energy conservation law and
by $z_x$ to obtain a momentum
conservation law. We prefer to show that the modified 
equation can be written 
in MS form and from this formulation 
obtain the associated local conservation laws via equations \rf{ecl}-\rf{mcl}.
Introducing the augmented variables 
\[
\tilde{z} = (z,z_t,z_{tt},z_x,z_{xx})^T,\quad
\tilde{S} = S + \frac{\Delta t^2}{24} z_{tt}^T\bM z_t + \frac{\Delta x^2}{24} z_{xx}^T\bK z_x,
\]
the modified equations \rf{MEQN} can be written in the MS form
\be
\mathbf{\tilde{M}}\tilde{z}_t +  \mathbf{\tilde{K}}\, \tilde{z}_{x} = 
\bgrad_{\tilde{z}} \tilde{S}(\tilde{z}),
\label{MMSEQN}
\ee
where  $\mathbf{\tilde{M},\, \tilde{K}}$
are the skew-symmetric matrices given by
\be
\mathbf{\tilde{M}} = \left(\ba{ccccc}
\bM  & \mb{0} & \frac{\Delta t^2}{24}\bM   & \mb{0} & \mb{0} \\ 
\mb{0} & - \frac{\Delta t^2}{24}\bM   & \mb{0} & \mb{0} & \mb{0} \\ 
\frac{\Delta t^2}{24}\bM   & \mb{0} & \mb{0} & \mb{0} & \mb{0} \\ 
\mb{0} & \mb{0} & \mb{0} & \mb{0} & \mb{0} \\
\mb{0} & \mb{0} & \mb{0} & \mb{0} & \mb{0}
\ea \right),
\quad  \mathbf{\tilde{K}} = \left(\ba{ccccc}
\bK   & \mb{0} & \mb{0} & \mb{0} & \frac{\Delta x^2}{24}\bK   \\
\mb{0} & \mb{0} & \mb{0} & \mb{0} & \mb{0} \\
\mb{0} & \mb{0} & \mb{0} & \mb{0} & \mb{0} \\
\mb{0} & \mb{0} & \mb{0} & - \frac{\Delta x^2}{24}\bK  & \mb{0} \\ 
\frac{\Delta x^2}{24}\bK   & \mb{0} & \mb{0} & \mb{0} & \mb{0} 
\ea \right)\nn.
\ee

The modified multisymplectic PDE can be used to derive the
modified LECL and LMCL.
Substituting $\tilde{z}$, $\mathbf{\tilde{M},\, \tilde{K}}$ and  $\tilde S$,  
into  (\ref{ecl}) and (\ref{mcl}), the modified LECL and LMCL 
are found to be, respectively,
\bea
\tilde{E}_t + \tilde{F}_x &=&
\left[ E+\frac{\Delta t^2}{24}z_{tt}^T\bM z_t +
\frac{\Delta x^2}{48}z_{xx}^T\bK z_x \right]_t +
\left[ F+\frac{\Delta x^2}{48}z_{xt}^T\bK z_x \right]_x = 0,\nn\\
\tilde{G}_x + \tilde{I}_t  &=&
\left[G+\frac{\Delta t^2}{48}z_{ttt}^T\bM z + 
\frac{\Delta x^2}{24}z_{xx} \bK z_x \right]_x\nn\\
&& +
\left[I+\frac{\Delta t^2}{48}\left(z_{xt}^T\bM z_t
+z_{tt}^T\bM z
+2 z^T\bM z_{xtt}\right)\right]_t = 0,\nn
\eea
where $E,\,F,\,G,$ and $I$ are given by equations \rf{nls_ecl}-\rf{nls_mcl}.
In the next section, we numerically verify that 
these modified local conservation laws 
are satisfied to higher order.

\section{Numerical Results}
For our numerical experiments we  consider the NLS equation 
with periodic boundary conditions, $u(x+L,t) = u(x,t)$.
We use initial data for a
 multi-phase quasi-periodic (in time) solution,
i.e., $u_0(x) = 0.5(1+ 0.1\cos\mu x),$
$\mu = 2\pi/L$, $L = 2\sqrt{2}\pi$. This initial data corresponds to
a multi-phase solution, near the plane wave, characterized by one excited mode.
We examine the performance of the centered cell box scheme 
(which we designate  as MS-CC) for varying mesh sizes and time steps.
The solution to equation \rf{ms-cc} is found by writing it in matrix form 
$\bA_- u^1 = \bA_+ u^0 + F(u^1,u^0)$
and using an iteration technique to solve for $u^1$.

\begin{figure}[htb]
\begin{center}
\includegraphics[height = 1.5in,width=2.25in]{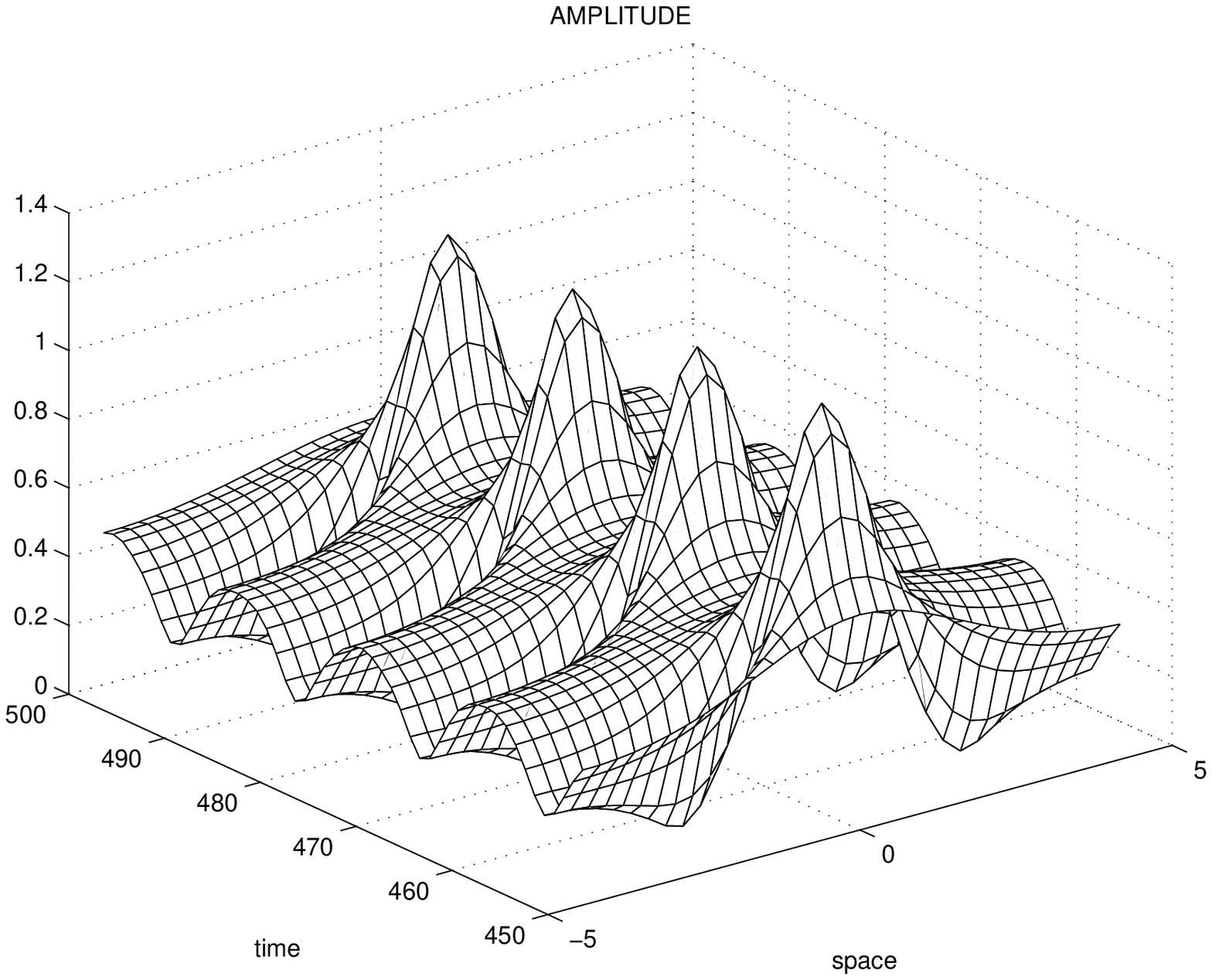}
\end{center}
\begin{center}
\includegraphics[height = 1.5in,width=2.25in]{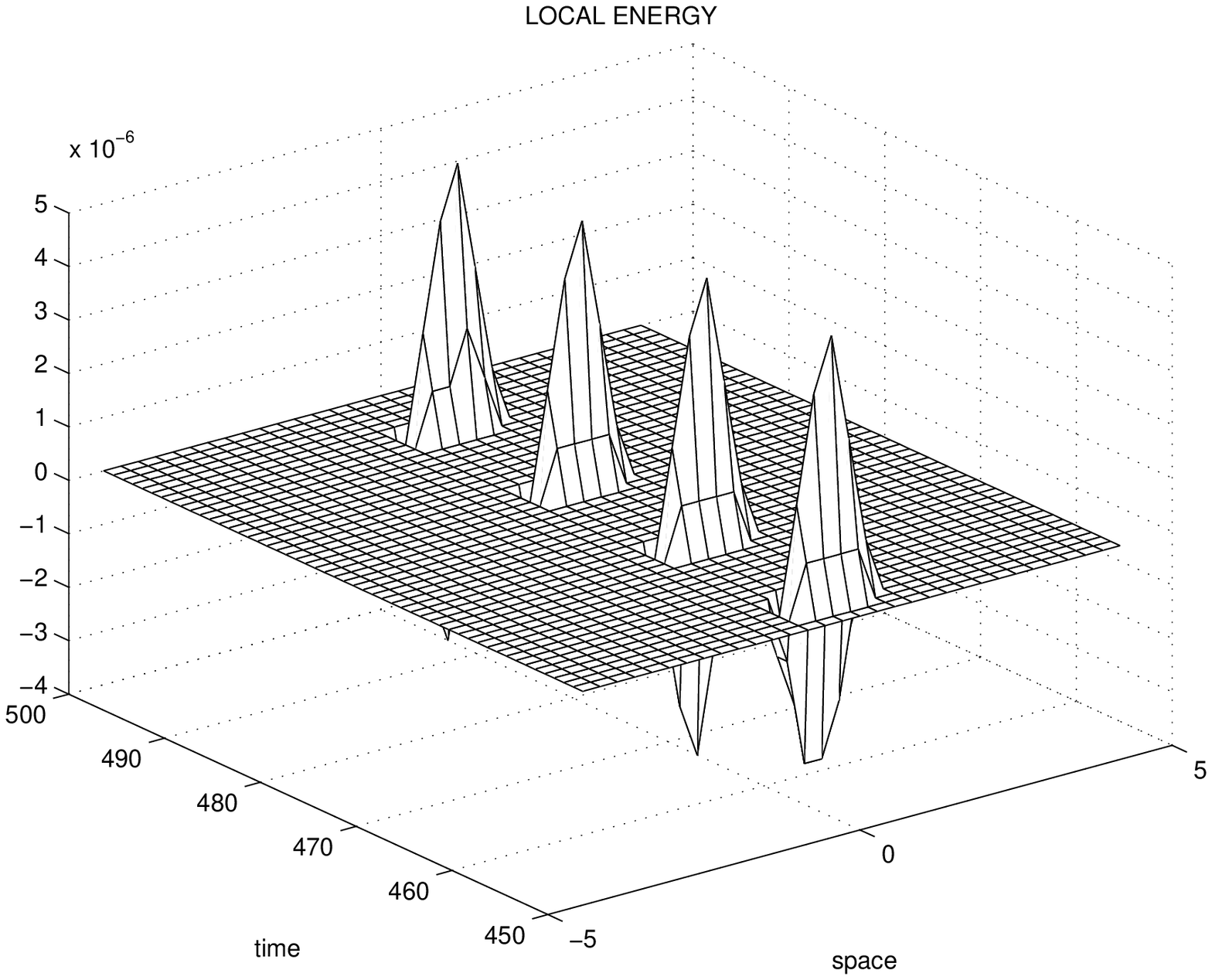}
\includegraphics[height = 1.5in,width=2.25in]{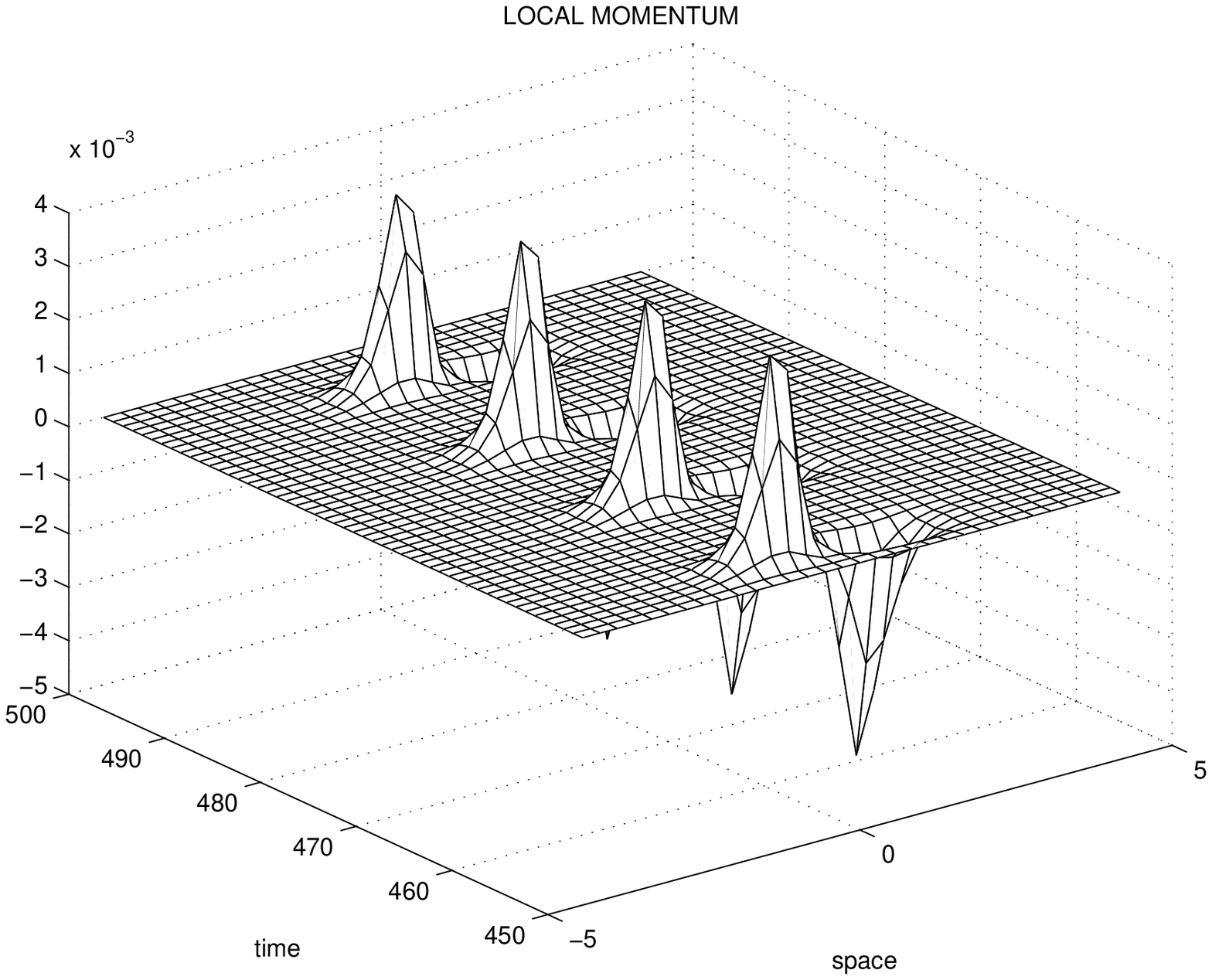}
\end{center}
\begin{center}
\includegraphics[height = 1.5in,width=2.25in]{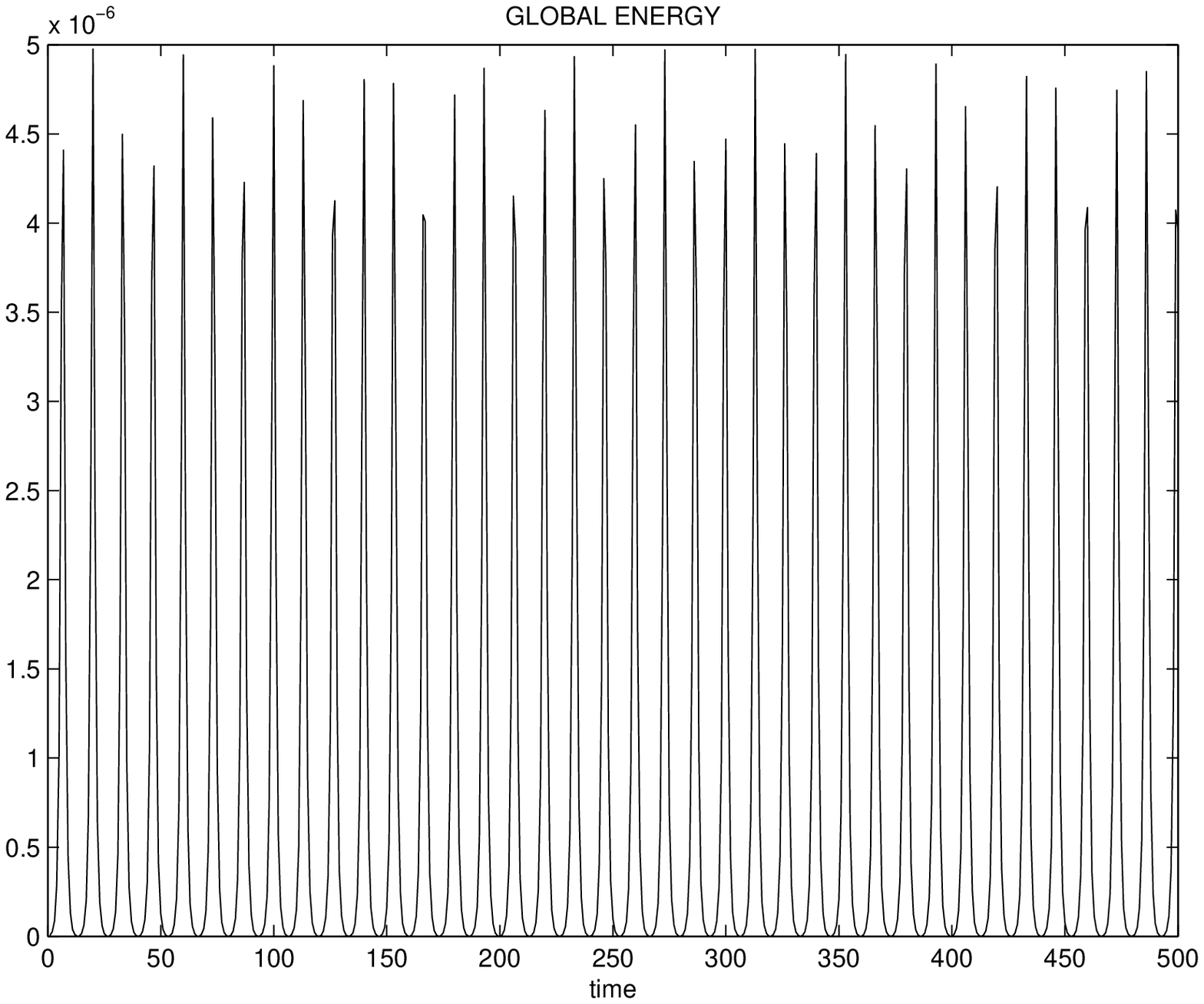}
\includegraphics[height = 1.5in,width=2.25in]{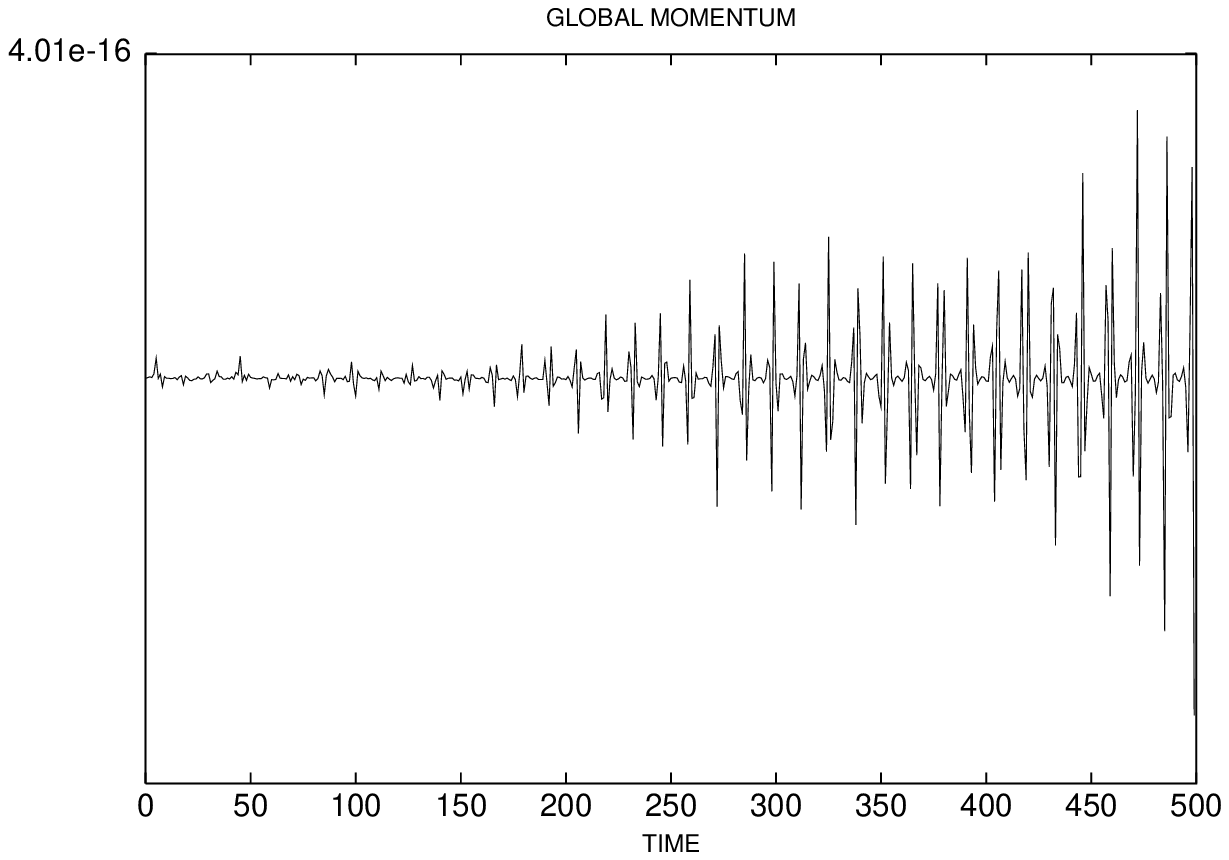}
\end{center}
\caption{The  MS scheme 
with $N = 64$ and  $t= 5\times 10^{-3}$,  $T = 500$:
a) surface,
b-c) error in the LECL  and LMCL  
d-e) error in the global energy and momentum.}
\label{fig2}
\end{figure}

The solution with $N = 64$ and  $dt= 5\times 10^{-3}$
for $450<t<500$ is shown in Figure~\ref{fig2}a.
The surface clearly exhibits the correct quasiperiodic 
behavior in time. In addition, we are interested in how well the
local and global conservation laws are satisfied.
To evaluate the local conservation laws, we use
midpoint discretizations of the form
\[
R_{1/2}^{1/2} = \frac{E_{1/2}^1 - E_{1/2}^0}{\Delta t} +
\frac{F^{1/2}_1 - F^{1/2}_0}{\Delta x}.
\]
In general, these residuals are not zero (Figures~\ref{fig2}b-c).
The errors in the local conservation laws,
the LECL and LMCL \rf{nls_ecl}-\rf{nls_mcl},
are $10^{-6}$ and $10^{-3}$, respectively, and are 
concentrated around the regions where there are steep gradients in
the solution.
If $S(z)$ were a quadratic functional of $z$, $S(z)=\half z^T\bA  z$,
with $\bA $ a symmetric matrix,
then the local conservation laws would be conserved exactly
\cite{br2}. In general, as in the present case,
the PDE is nonlinear and 
$S(z)$ is not a quadratic functional. Therefore, the local
energy and momentum conservation laws will not be preserved
exactly. However the numerical experiments show that 
the local conservation laws are preserved very well over long times.
In addition to  resolving the 
LECL and LMCL very well, the MS scheme 
preserves the global errors extremely well.
The error in the global energy  oscillates in a bounded
fashion, as expected  of 
a symplectic integrator (Figure~\ref{fig2}d) while
the error in the global momentum (Figure~\ref{fig2}e)
and the norm (not shown) are conserved
{\em exactly \/} (up to the error criterion of $10^{-14}$ in the solver)
since they are quadratic invariants.

The maximum error in the LECL and LMCL 
and in the global energy and momentum
for the MS scheme are provided 
in Table 1 for varying mesh sizes and  time steps.
The error in the  LECL depends only on the timestep $t$ and 
is second order, while the 
error in the LMCL depends
only on the spatial mesh size $N$ and 
is second order.
\begin{table}[htb]
\begin{tabular*}{\textwidth}{@{\extracolsep{\fill}}lcccccc}
\hline
N & 32 & 32 & 32 & 64 & 64 & 64 \\
$t$ & 2.0E-02 & 1.0E-02 & 5.0E-03 & 2.0E-02 & 1.0E-02 & 5.0E-03  \\ \hline
LE & 6.0E-05 & 1.5E-05 & 4.0E-06 & 8.0E-05 & 2.0E-05 & 5.0E-06 \\ \hline
LM & 1.7E-02 & 1.7E-02 & 1.7E-02 & 4.8E-03 & 4.8E-03 & 4.8E-03 \\ \hline
GE & 7.3E-05 & 2.0E-05 & 5.0E-06 & 7.6E-05 & 2.2E-05 & 5.0E-06 \\ \hline
GM & 1.2E-13 & 2.5E-14 & 2.0E-13 & 1.3E-13 & 1.0E-13 & 4.5E-13 \\ \hline
\end{tabular*}
\caption{The absolute maximum error in the local energy and momentum 
and the global energy and momentum obtained using the MS scheme
MS, with T = $500$.}
\end{table}

We next examine whether the modified local conservation laws
obtained using the MS-CC discretization of the NLS 
are preserved to a higher order than the original local
conservation laws. 
Since our solution is quasiperiodic, we compute the solution
for $0<t<T$, where $T$ is chosen
to include a characteristic cycle. From Figure~\ref{fig2}, $T=10$
is sufficient. Since the ECL is independent of $\Delta x$ (see Table~1),
for a fixed $N$, we let $\Delta t\rightarrow 0$.
That is, start with $\Delta t_0 = \Delta x$
and let $\Delta t = \Delta t_0/2^k,\,k=0,1,...,6$.

We compute the LECL and the modified LECL at each time step using centered
approximations of the derivatives of sufficiently high order so as  not to
affect the order of the MS-CC discretization of the residuals.
Figure~\rf{fig3} shows the loglog plot of the maximum error as a function
of the timestep for the LECL and the  modified LECL. 
\begin{figure}[htb]
\begin{center}
\includegraphics[height = 3in,width=3in]{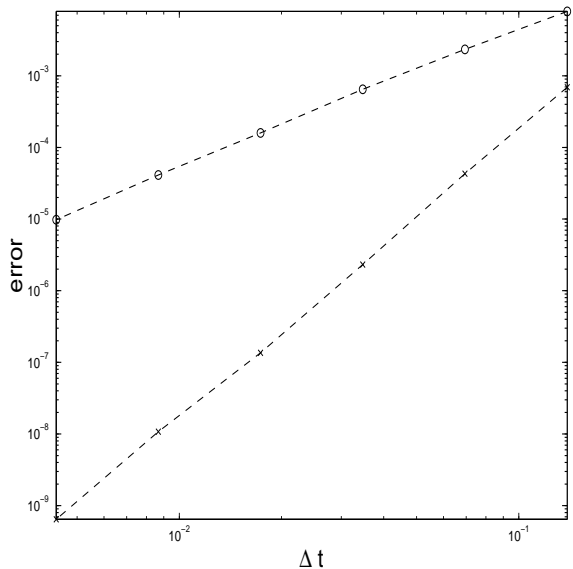}
\end{center}
\caption{Loglog plot of the error against $\Delta t$ 
for the original ECL (o - -), and the MECL (x - -).}
\label{fig3}
\end{figure}
Clearly we can see that while the
LECL is satisfied to 2nd order, the modified LECL is satisfied to 4th order.
Verification of higher preservation of the LMCM as a function of
the mesh size is similar.

\section*{Acknowledgements}
This work was partially supported by the
NSF, grant number DMS-0204714.


\begin{thebibliography}{99}

\bibitem{mcl03} {U. Ascher and R. McLachlan}, Multisymplectic box schemes 
and the Korteweg-de Vries equation, preprint 2003.

\bibitem{br2} {T.J. Bridges and S. Reich},
Physics Letters A,  \textbf{284}, 184-193 (2001).

\bibitem{hlu97} {E. Hairer and Ch. Lubich,} Numer. Math., 
\textbf{76},  441 (1997).

\bibitem{htext} {E. Hairer, Ch. Lubich and G. Wanner},
Geometric Numerical Integration, Springer Verlag, Berlin, 2002.

\bibitem{iskasc} {A.L. Islas, D.A. Karpeev and C.M. Schober},
J. of Comp. Phys.  \textbf{173}, 116--148 (2001).

\bibitem{issc03} {A.L. Islas and C.M. Schober,}
Fut. Gen. Comp. Sys,  \textbf{19}, 403 (2003).

\bibitem{issc03b} {A.L. Islas and C.M. Schober},  On the preservation of 
phase space structure under multisymplectic discretization, accepted 
J. of Comp. Phys.  2003.

\bibitem{mars2} J.E. Marsden and S. Shkoller,
Math. Proc. Camb. Phil. Soc. \textbf{125}, 553--575 (1999).

\bibitem{moore03b} {B. Moore and S. Reich,} Fut. Gen. Comp. Sys., \textbf{19},
395 (2003).

\bibitem{moore03} {B. Moore and S. Reich,} Num. Mathematik, \textbf{95}, 625
(2003).

\bibitem{reich1} {S. Reich},
J. of Comp. Phys.  \textbf{157}, 473--499 (2000).

\bibitem{pdetext} {J.W. Thomas,} Numerical Partial Differential Equations,
Springer Verlag, New York, 1995.

\end{thebibliography}
\end{document}